# Absolute frequency measurement of the $^{171}$Yb optical lattice clock at KRISS using TAI for over a year

Huidong Kim, Myoung-Sun Heo, Chang Yong Park, Dai-Hyuk Yu, and Won-Kyu Lee*

Korea Research Institute of Standards and Science, Daejeon 34113, South Korea

*Corresponding author e-mail: oneqlee@kriss.re.kr

## Abstract

We report a measurement of the absolute frequency of the $^1S_0$-$^3P_0$ transition in the $^{171}$Yb optical lattice clock at KRISS (KRISS-Yb1) for 14 months, which was referenced to the SI second by primary and secondary standards worldwide via TAI (International Atomic Time). The determined absolute frequency is 518 295 836 590 863.75(14) Hz with the relative frequency uncertainty of $2.6\times10^{-16}$, which agrees well with other reports. This result is expected to contribute to the future update of the CIPM recommendation frequency of the secondary frequency standards. (In Corregendum at the end of this article, the determined absolute frequency is corrected by using the new height measurement )

## 1. Introduction

Since the SI unit of time (the second) was defined using a ground-state hyperfine transition frequency of the $^{133}$Cs atom in 1967 [1], its practical realization by primary frequency standards has been improved, reaching the low $10^{-16}$ uncertainty level [2-4]. While there have been no further significant breakthroughs in the frequency uncertainty of Cs primary frequency standards over the last two decades, the performances of optical clocks began to surpass those of the primary standards since around 2005, and the number of the optical clocks included in the secondary representation of the second is increasing [5]. Several optical clocks have already achieved the frequency uncertainty around $1\times10^{-18}$ level [6-10], and the roadmap for the redefinition of the second is seriously being discussed [5, 11]. One of the required conditions for this redefinition is the continuity with the current definition based on Cs. Thus, accumulating more results of the independent absolute frequency measurements of the optical clocks is highly desirable and examining the consistency between the measurement values is very important [5].

Optical lattice clocks using Yb atoms are among the best candidates for the redefinition of the second and are being investigated worldwide [12-24]. The absolute frequency measurements of the $^1S_0$-$^3P_0$ transition of $^{171}$Yb have been performed many times by laboratories worldwide by using TAI (International Atomic Time) [13, 25-31], by using a local Cs fountain clock [20, 32], or by using the frequency ratio measurement between different atomic species [17, 33-42].

We performed a new absolute frequency measurement of an Yb optical lattice clock (KRISS-Yb1) for 14 months (from January 2020 to February 2021) with a total measurement time of 400.5 h. This was referenced to the SI second realized by primary or secondary frequency standards worldwide reported to the International Bureau of Weights and Measures (BIPM) during the measurement campaign.

The organization of this paper is as follows. In Section 2, we outline the KRISS-Yb1 optical lattice clock apparatus and the frequency measurement scheme that links to the SI second. In Section 3, we describe the frequency shift and its uncertainty in each step of the measurement scheme. Finally, We report the absolute frequency measurement result during this 14-month-long campaign and compare it with other reports in Section 4.

## 2. Experimental methods and apparatus

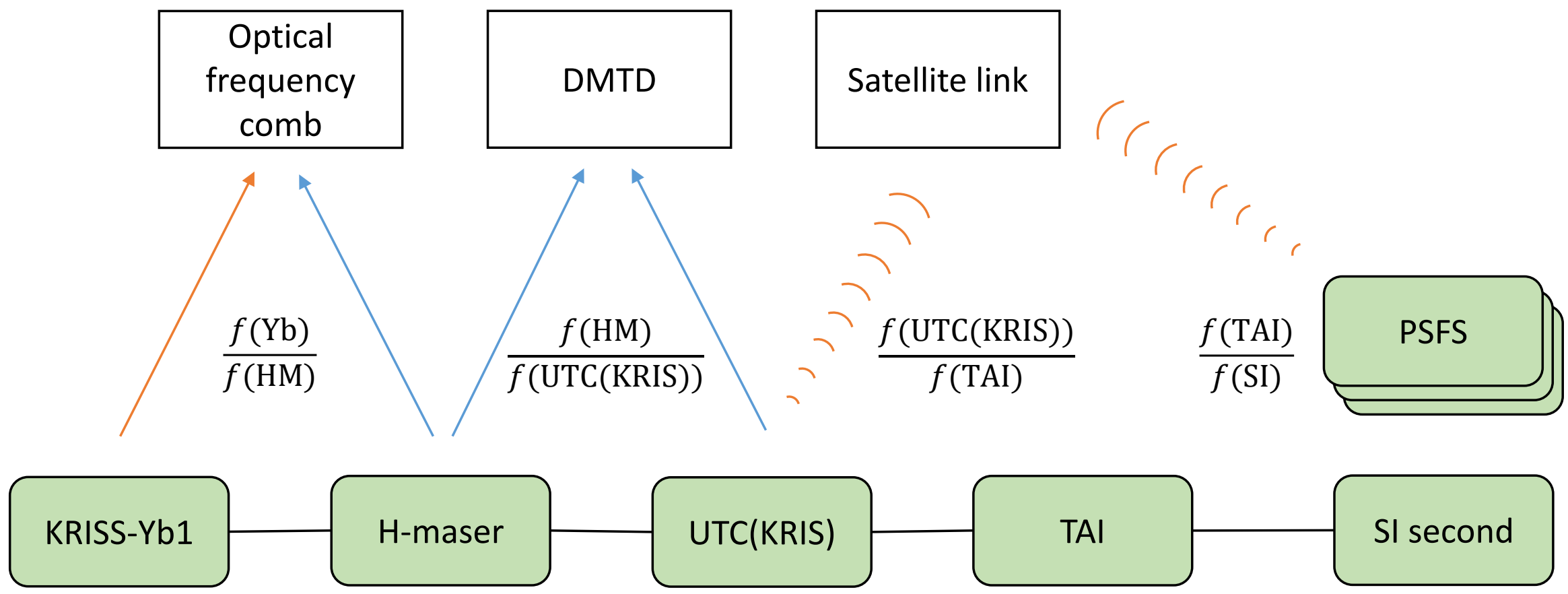

**Figure 1.** Frequency chain schematic diagram for the absolute frequency measurement relative to the SI second (Terrestrial time, TT). DMTD: multi-channel dual-mixer time difference, PSFS: primary or secondary frequency standards

We measured the absolute frequency of KRISS-Yb1 with a link to TAI using a hydrogen maser (HM) as a flywheel. The schematic diagram of the frequency measurement chain is shown in figure 1. The frequency of the HM was continuously measured locally relative to UTC(KRIS). UTC(KRIS) is compared with TAI by satellite links. These comparison results are announced in the Circular T [43] which is published by the BIPM monthly. The BIPM realizes TAI at 5-day intervals and its deviation from the SI second is also published in the Circular T. The absolute frequency measurement chain in this research is given by

$$\frac{f(\mathrm{Yb})}{f(\mathrm{SI})} = \frac{f(\mathrm{Yb})_{T(up)}}{f(\mathrm{HM})_{T(up)}} \cdot \frac{f(\mathrm{HM})_{T(up)}}{f(\mathrm{HM})_{T(5d)}} \cdot \frac{f(\mathrm{HM})_{T(5d)}}{f(\mathrm{UTC(KRIS)})_{T(5d)}} \cdot \frac{f(\mathrm{UTC(KRIS)})_{T(5d)}}{f(\mathrm{TAI})_{T(5d)}} \cdot \frac{f(\mathrm{TAI})_{T(5d)}}{f(\mathrm{TAI})_{T(m)}} \cdot \frac{f(\mathrm{TAI})_{T(m)}}{f(\mathrm{SI})_{T(m)}} , \tag{1}$$

where $f(\mathrm{Yb})/f(\mathrm{SI})$ is the absolute frequency of the Yb optical lattice clock, $f(\mathrm{SI}) = 1$ Hz, $f(\mathrm{Yb})_{T(up)}/f(\mathrm{HM})_{T(up)}$ is the measured frequency of the Yb optical lattice clock relative to the HM during the intermittent uptime of the optical lattice clock, $f(\mathrm{HM})_{T(up)}/f(\mathrm{HM})_{T(5d)}$ reflects the frequency drift and the dead-time uncertainty of the HM flywheel in the 5-day grid, $f(\mathrm{HM})_{T(5d)}/f(\mathrm{UTC(KRIS)})_{T(5d)}$ is the frequency of the HM relative to UTC(KRIS) during the 5-day grid, $f(\mathrm{UTC(KRIS)})_{T(5d)}/f(\mathrm{TAI})_{T(5d)}$ is the frequency of UTC(KRIS) relative to TAI, $f(\mathrm{TAI})_{T(5d)}/f(\mathrm{TAI})_{T(m)}$ reflects the dead-time uncertainty of 5-day grid TAI during the corresponding month interval (25, 30, or 35 days depending on the month), and $f(\mathrm{TAI})_{T(m)}/f(\mathrm{SI})_{T(m)}$ is the monthly frequency deviation of TAI from the SI second. The experimental methods and apparatus in each step of equation (1) are described in this section.

### 2.1. Improvement of Yb optical clock system

Our first Yb optical clock (KRISS-Yb1) has been operated for more than ten years and we reported twice the results of the absolute frequency measurements via TAI in 2013 [27] and 2017 [28], which contributed to the value of the recommended frequency for $^{171}$Yb as a secondary representation of the second [5]. Furthermore, we identified the relation between the pulse area of the Rabi spectroscopy and the density shift [15]. There was a remote frequency ratio measurement between KRISS-Yb1 and NICT-Sr1 via an advanced satellite-based frequency transfer method (two-way carrier-phase; TWCP) [40]. There have been significant system improvements between each report. In this section, we summarize the Yb1 system improvements after the previous reports [27, 28].

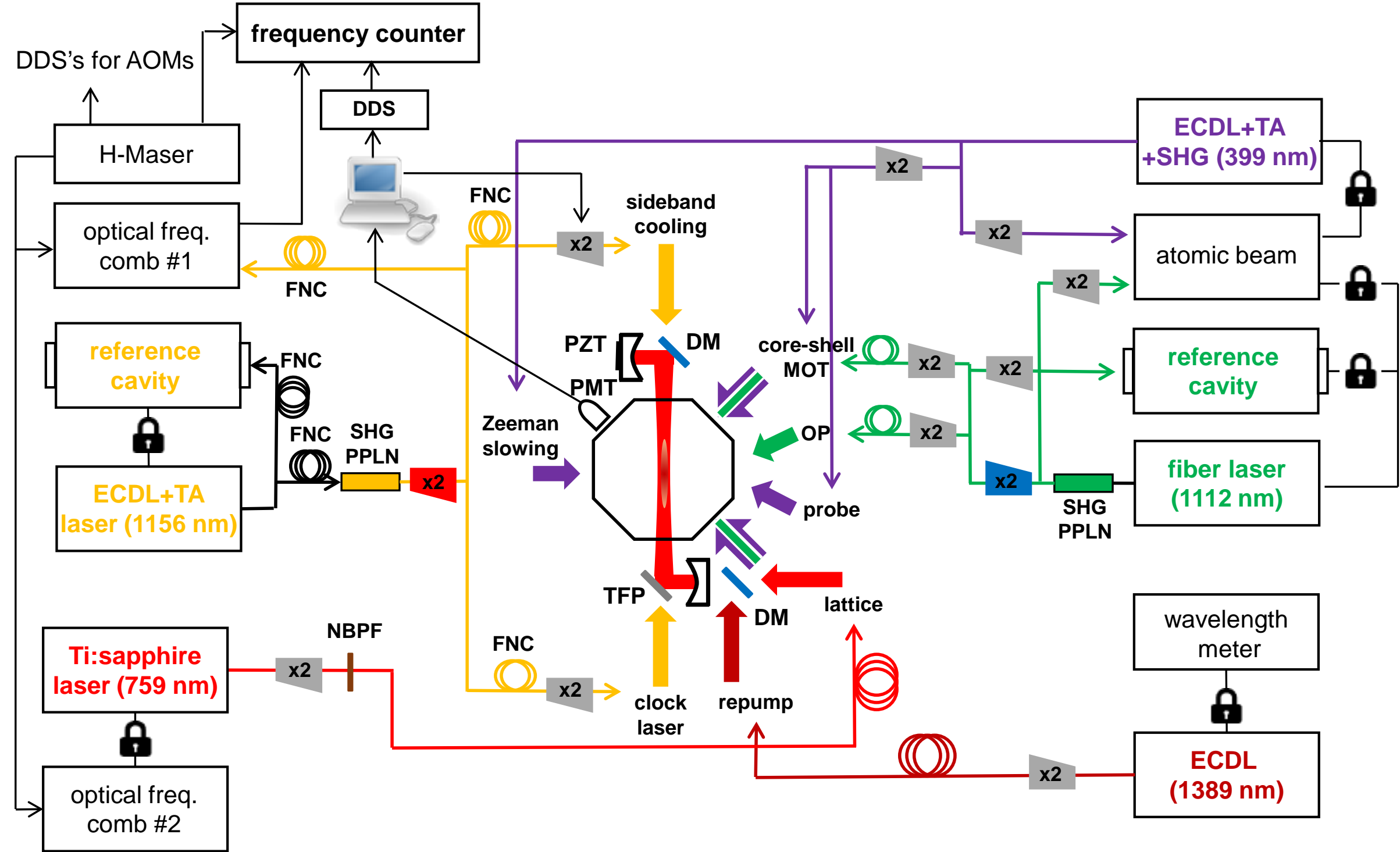

**Figure 2.** Yb optical clock schematics. DDS: direct digital synthesizer, ECDL: external cavity diode laser, TA: tapered amplifier, FNC: fiber noise cancellation, DM: dichroic mirror, OP: optical pumping, TFP: thin film polarizer, SHG: second-harmonic generation, PPLN: periodically-poled lithium niobate, NBPF: narrow bandpass filter, x2: double-pass AOM (the red AOM is for the frequency drift compensation and the blue AOM is for the fast feedback of the frequency stabilization), PZT: piezoelectric transducer, PMT: photo-multiplier tube.

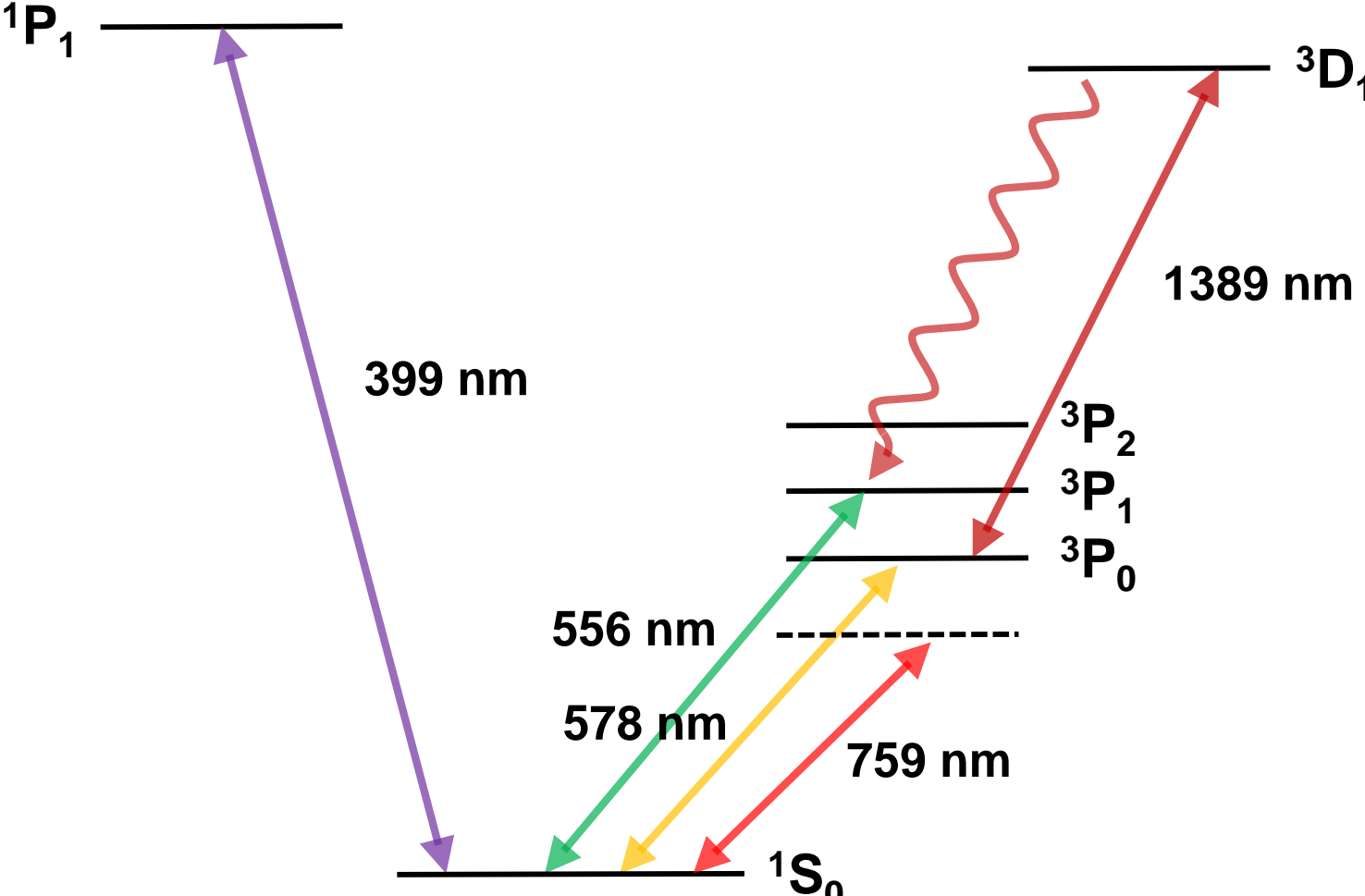

**Figure 3.** Optical transitions between energy levels used in the Yb optical clock.

The overall configuration of the improved apparatus for this measurement is shown in figure 2. The most significant feature of improvements is the power build-up cavity (finesse ~ 220) for the optical lattice, formed with two cavity mirrors and two folding mirrors. One of the two folding mirrors is a thin film polarizer (TFP), which maintains the polarization of the intracavity light parallel to the quantization axis. The length of the lattice cavity was stabilized by the Hänsch–Couillaud method [44]. We monitored the intracavity power of the lattice laser using the transmitted light through the TFP. We combined the 578 nm laser for Rabi spectroscopy parallel to the lattice light through the same TFP. The other folding mirror is a dichroic mirror (DM), which combines the 578 nm laser for the sideband cooling.

For the Yb optical clock operation, we need five different light sources with a wide wavelength range from near UV to IR, shown in figure 3. The first cooling laser at the $^1S_0$-$^1P_1$ singlet transition (399 nm) was obtained from a commercial laser system composed of an external cavity diode laser (ECDL), a tapered amplifier and a second harmonic generation (SHG) cavity (TA-SHG pro, TOPTICA). The output power of the 399 nm laser system was about 1 W, but the power of 100 mW in total was sufficient for the Zeeman slowing, the laser cooling, the frequency stabilization, and the probe beam. The 399 nm laser frequency was stabilized by the FM spectroscopy technique using a fluorescence signal from an Yb atomic beam machine. The frequency offset of -520 MHz for the Zeeman slowing was applied by a double-pass acousto-optic modulator (AOM) on the path to the Yb atomic beam machine. We tuned the laser frequencies for the singlet magneto-optical trap (blue MOT) and the probe beam using another double-pass AOM. The laser frequency for the blue MOT was red-detuned by 24 MHz from the resonance.

The second cooling laser at the $^1S_0$-$^3P_1$ triplet transition (556 nm) was obtained by the SHG of a commercial Yb-doped fiber laser system at 1112 nm using a fiber-coupled ridge-type waveguide periodically poled lithium niobate (WG-PPLN, NTT). The total power of 170 mW at 556 nm was used for the laser cooling, the frequency stabilization, and the spin-polarization in KRISS-Yb1. The linewidth of the 556 nm laser was less than the natural linewidth of the triplet transition (182 kHz) after frequency stabilization using both a high-finesse ultra-low expansion glass (ULE) cavity and an Yb atomic beam. As a result, we could maintain the frequency locking for more than a week and make the long-term operation of the KRISS-Yb1 [45, 46]. The laser frequencies were tuned for the triplet MOT (green MOT) and for the spin-polarization by using respective double-pass AOMs. These laser beams were sent to the Yb spectroscopy chamber using polarization-maintaining fibers (PMFs).

The lattice laser at 759 nm for the vertical one-dimensional optical lattice was obtained from a commercial Ti:sapphire laser system (SolsTiS, M Squared Lasers). A small part of the output power was delivered via a PMF to an optical frequency comb (Menlo Systems) for the frequency stabilization and to a visible wavelength meter (HighFinesse) for the frequency monitor. We phase-locked the 759 nm laser to the optical frequency comb, which is referenced to a HM, and the resulting laser linewidth was about 100 kHz. The remainig laser beam (~ 1 W) was sent through a double-pass AOM for the power stabilization and frequency tuning, and then was transferred for the optical lattice by a 2-m-long large-mode-area PMF (LMA-PM-5, NKT). We controlled the lattice trap depth ($U_0$) between 48 $E_R$ and 1000 $E_R$, where $E_R$ is the recoil energy of a lattice photon, using a power servo and an amplitude modulation by the AOM. We inserted a narrow bandpass filter (Alluxa) with a FWHM (full width at half maximum) of 1 nm before the fiber coupling to reduce the residual ac Stark shift caused by the amplified spontaneous emission (ASE) of the lattice laser. It is noted that the ASE effect was further reduced by the build-up cavity for the optical lattice.

The repumping laser at 1389 nm was obtained from a commercial ECDL system (DL pro, Toptica). The output power of about 10 mW at 1389 nm was used for the sideband cooling [17] and the electron-shelving detection technique [47]. The laser frequency was stabilized using an IR wavelength meter (HighFinesse) [48]. The laser output was transferred to the Yb spectroscopy chamber by a PMF. We tuned the laser frequency by a double-pass AOM for the sideband cooling and for the detection of the excited atom of the clock transition. We combined the 1389 nm laser with the 759 nm optical lattice light using a dichroic mirror.

The previous clock laser system at 578 nm using the sum frequency generation [49] has been replaced by a new one using the SHG of a laser source at 1156 nm [50, 51] since August 2019. This new clock laser system is based on a 30-cm-long ULE cavity with $4.8\times10^{-17}$ thermal noise limit at room temperature [52] adopting fused-silica mirror substrates and crystalline mirror coatings [53] to reduce the thermal noise. The temperature of the cavity was stabilized at 25.013°C, zero crossing temperature of the coefficient of thermal expansion of the ULE spacer. The frequency stability was measured to be $5.6\times10^{-16}$ at 1 s, currently limited by unidentified technical noise sources, but is still capable of making a shorter averaging time required in the Yb clock uncertainty evaluation than before. The output power of the 1156 nm laser (an ECDL followed by a tapered amplifier) was about 500 mW. The power of the SHG was more than 20 mW using a ridge-type WG-PPLN. The SHG output at 578 nm was distributed to an optical frequency comb and the Yb spectroscopy chamber using PMFs with the fiber noise cancelled for each path [54]. The linear drift of the clock laser at 578 nm due to ULE material aging was 0.076 Hz/s initially and decreased to 0.036 Hz/s currently. This linear drift was removed using a double-pass AOM, and the residual frequency drift was kept below 1 mHz/s by updating the frequency drift rate of the AOM once in several weeks.

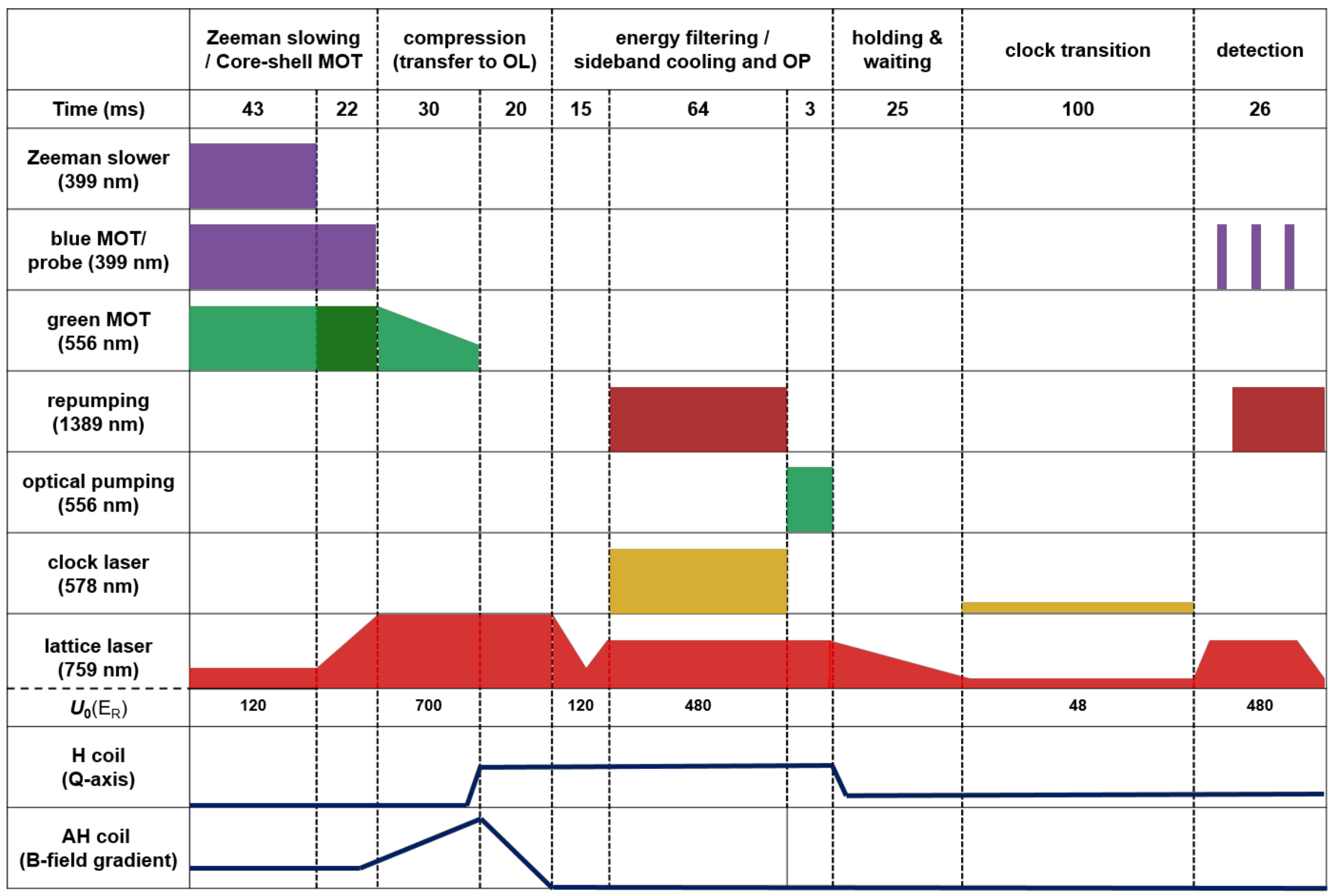

**Figure 4.** Experimental sequence of KRISS-Yb1. OL: optical lattice, OP: optical pumping for the spin-polarization of the initial state of trapped Yb atoms, H coil: Helmholtz coil, AH coil: anti-Helmholtz coil, Q-axis: quantization axis. In the dark green region, the frequency of the 556 nm laser was modulated for better transfer of atoms to the green MOT.

Our experimental sequence consists of six stages shown in figure 4. Firstly, we cooled and trapped Yb atoms from an atomic beam oven at about 300°C using a Zeeman slower and a double-stage MOT. In order to improve the trap loading rate and the steady-state atom number of the MOT, we adopted the core-shell MOT technique [55] using a circular mask with a diameter of 5 mm. As a result, the loading time and the first cooling laser power could be reduced by 1/3 and 1/5, respectively, compared to the conventional two-stage method previously used. After the end of Zeeman slowing, we modulated the frequency of the second cooling laser (556 nm) sinusoidally at 29 kHz with a frequency deviation of 2 MHz using an AOM to improve the transfer efficiency from the blue MOT to the green MOT (dark green region in figure 4). At the same time, the optical lattice depth was ramped up to 700 $E_R$ to maximize the number of atoms trapped in the lattice before the additional cooling processes (the energy filtering and the sideband cooling) which are to be described below.

In the second sequence, to load as many atoms to the optical lattice as possible, we compressed the green MOT by the magnetic field gradient ramp-up from 7 G/cm to 20 G/cm; furthermore, the power of the second cooling laser was ramped down, and the detuning was tuned close to the resonance of the triplet transition. And then, while the atoms were held in the trap, we ramped down and turned off the magnetic field gradient within 20 ms. To reduce the axial temperature of trapped atoms, we applied additional cooling processes using the energy filtering [15, 56] and the sideband cooling [57, 58]. For the energy filtering, the trap depth was ramped down to 120 $E_R$ in 10 ms and ramped up to 480 $E_R$ in 5 ms. And then, for the sideband cooling, Yb atoms were excited from an axial vibrational state $n$ in $^1S_0$ to a state $n-1$ in $^3P_0$ by a laser tuned near the low-frequency edge of the red sideband of the clock transition at 578 nm. We checked the atomic temperature using the sideband spectrum, which was obtained by applying a clock laser of Rabi frequency of ~2 kHz. After sideband cooling, spin-polarization to $m_F = +1/2$ or -1/2 in the ground state $^1S_0$ was carried out using the near-resonant light to the 556 nm triplet transition under the quantization magnetic field of 4.7 G.

Before the clock transition, we adiabatically ramped down the trap depth to the target clock operation depth of 48 $E_R$. In this case, the average axial vibrational state $\bar{n}$ was usually kept around 0.15. And then, the clock transition ($^1S_0$–$^3P_0$) was interrogated with a clock laser at 578 nm using a square $\pi$-pulse with a duration of 100 ms in the presence of a 0.75 G magnetic field, which induced the Zeeman splitting of 300 Hz between the two $\pi$-transitions.

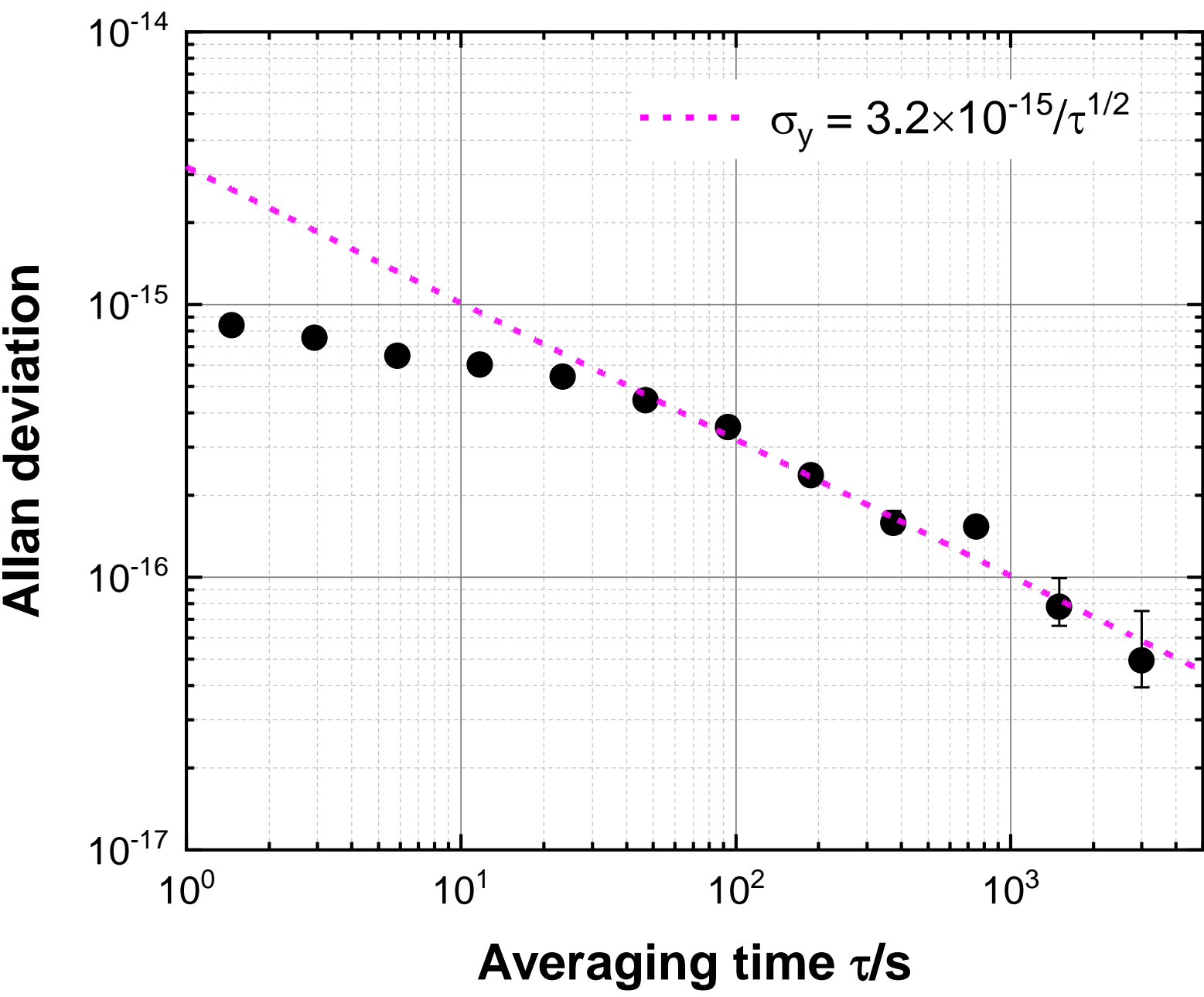

**Figure 5.** Stability of KRISS-Yb1 in terms of overlapping Allan deviation of the frequency difference between the two interleaved measurements.

With the improved clock laser system, it was expected that the short-term instability of KRISS-Yb1 could be greatly reduced. However, this was limited by the shot-to-shot magnetic field fluctuation induced by the optical table and the vacuum chamber which were not made of highly non-magnetic material. As a result, the stability of KRISS-Yb1 was $3.2 \times 10^{-15}/\sqrt{\tau}$, as shown in figure 5. To measure the long-lived upper clock state ($^3P_0$) population, we used a normalized electron-shelving technique [47]. After the clock transition, we probed the atoms in the ground state using the 399 nm probe light until all atoms scattered out of the lattice while collecting the fluorescence signal using a photo-multiplier tube (PMT). After applying the repumping laser at 1389 nm, the second probe light was used to detect atoms in the excited state. And finally, we applied the third probe light to measure the background scattering. Using these three successive fluorescence signals, we could calculate the normalized atomic excitation fraction.

A single cycle of the clock operation, which took about 1.5 s, consisted of two pairs of experimental sequences in figure 4 (at two frequency detuning values with the opposite sign around the clock transition for each of the spin polarization to either $m_F = +1/2$ or $m_F = -1/2$). Each pair provided the frequency offset from the center of each clock transition, and the combined offset values cancelled out the first-order Zeeman shift. We obtained the frequency correction signal using a digital proportional-integral-differential (PID) filter, and applied this signal to an AOM for the clock laser. The frequency of the DDS driving this AOM was measured by a frequency counter. Although the first-order Zeeman shift was cancelled out using the above process, long-term variation of an ambient magnetic field due to environmental changes was inevitable, resulting in the variation of the quadratic Zeeman effect. Thus, the current through the Helmholtz coil for the quantization magnetic field was modulated with a time constant of 150 seconds so that we could maintain the measured Zeeman splitting of 300 Hz between the two $\pi$-transition peaks.

### 2.2. Optical frequency comb and frequency counter

Two similar erbium fiber laser frequency combs were used in this experiment. One was used for the optical phase-lock of the 759 nm lattice laser, and the other was used for the frequency measurement of the 578 nm clock laser. The repetition rates of the frequency combs (~250 MHz) were stabilized using a HM as a frequency reference. The fourth harmonic of the repetition rate was mixed with an output of a dielectric resonator oscillator at 980 MHz, which was phase-locked to the HM reference signal, to produce an intermediate frequency (IF) signal at about 20 MHz, and then the IF signal was phase locked to the DDS output at a fixed frequency near 20 MHz. All the RF frequency synthesizers and frequency counters were also referenced to the same HM. The beat note frequency between the clock laser and the adjacent optical frequency comb mode was measured using a dead-time-free multi-channel frequency counter (K+K Messtechnik FXE) in Π-mode [59] with a gate time of 1 s. We measured this beat note frequency using another redundant channel of the counter to remove the data with cycle slips by excluding the data with the difference of the two measurements greater than 1 Hz. The frequency of a direct digital synthesizer (DDS) used in the frequency lock to the Yb atom was also measured simultaneously using a separate channel of the same frequency counter for the absolute frequency measurement.

### 2.3. H-maser and DMTD

A HM named H5628 (BIPM code 1405628) was used as a frequency reference for the measurements in the year 2020. However, as H5628 showed large unpredictable frequency excursions departing from linear drift, we changed the frequency reference to another HM named H5626 (BIPM code 1405626) since January 2021. All the data obtained in 2020 could be processed to be referenced to H5626 using the phase comparison data from a multi-channel dual-mixer time difference (DMTD) measurement system. The frequency difference between H5626 and UTC(KRIS) could also be calculated using the same DMTD data. The frequency drift of H5626 relative to TT during the absolute frequency measurement campaign is shown in figure 6 with the average drift value of $-4.5\times10^{-17}$/d. The frequency stability of H5626 is shown in figure 7, which was obtained by comparisons with two other HMs (by three-cornered hat method, blue dotted line), with the Yb optical lattice clock (pink circles), with a Cs fountain clock (KRISS-F1, blue triangles), and with TT (green squares). We conservatively estimate the noise of H5626 as the square sum of white phase noise $1.0\times10^{-13}$ $(\tau/s)^{-1}$, white frequency noise $1.2\times10^{-13}$ $(\tau/s)^{-1/2}$, and flicker frequency noise $3\times10^{-16}$. This noise model is shown as an orange solid line in figure 7 and covers well all the measured frequency instability including small bump around 4000 s.

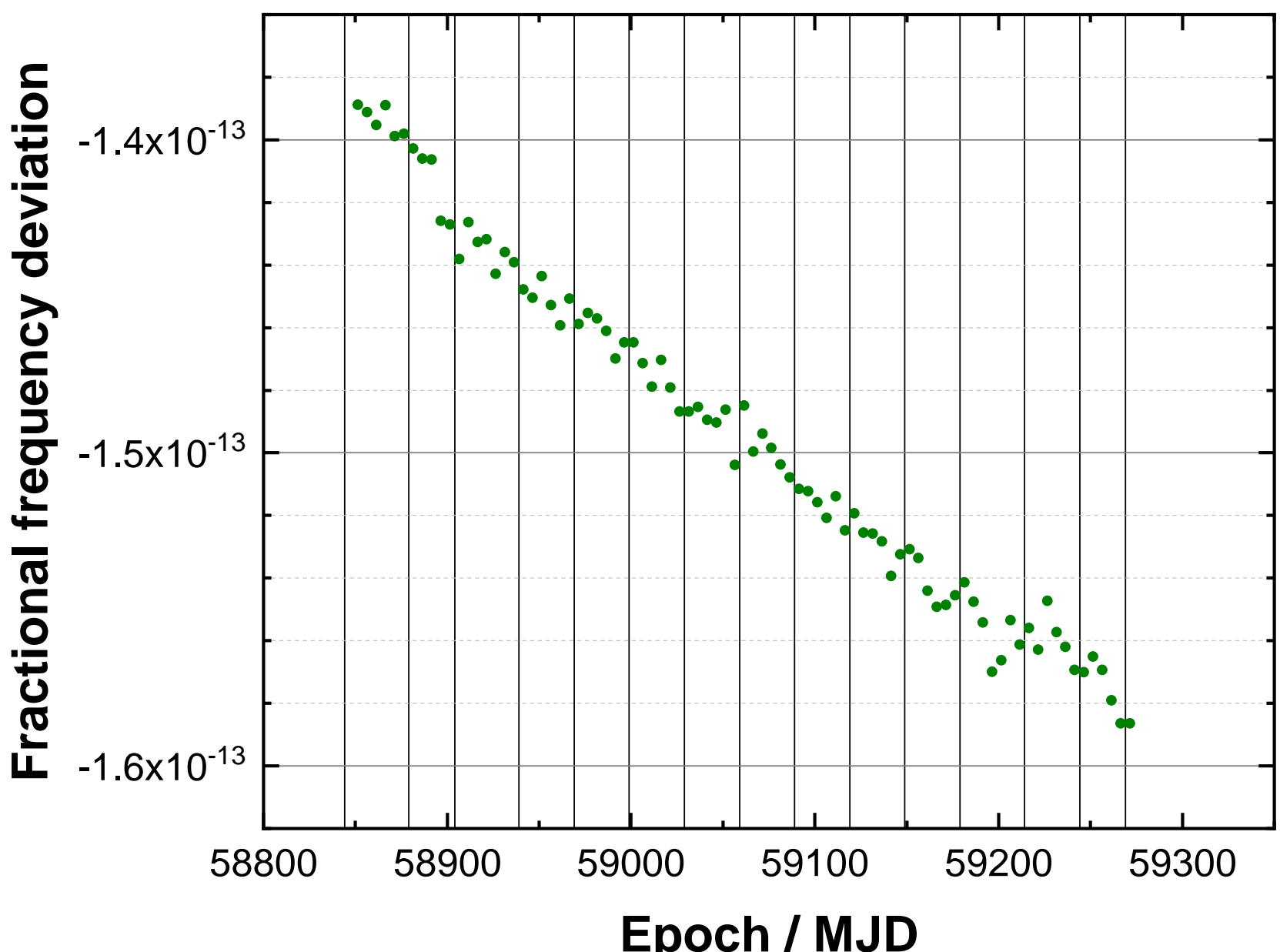

**Figure 6.** Fractional frequency deviation of the HM (H5626) measured by TT during the absolute frequency measurement campaign (MJD 58844 ~ 59269) with the average drift value of $-4.5\times10^{-17}$/d. The vertical solid lines represent the separated interval for each Circular T. The drift rate and its uncertainty for each month-grid of the Circular T was determined by a linear fit for 3 months (including the previous and the next month).

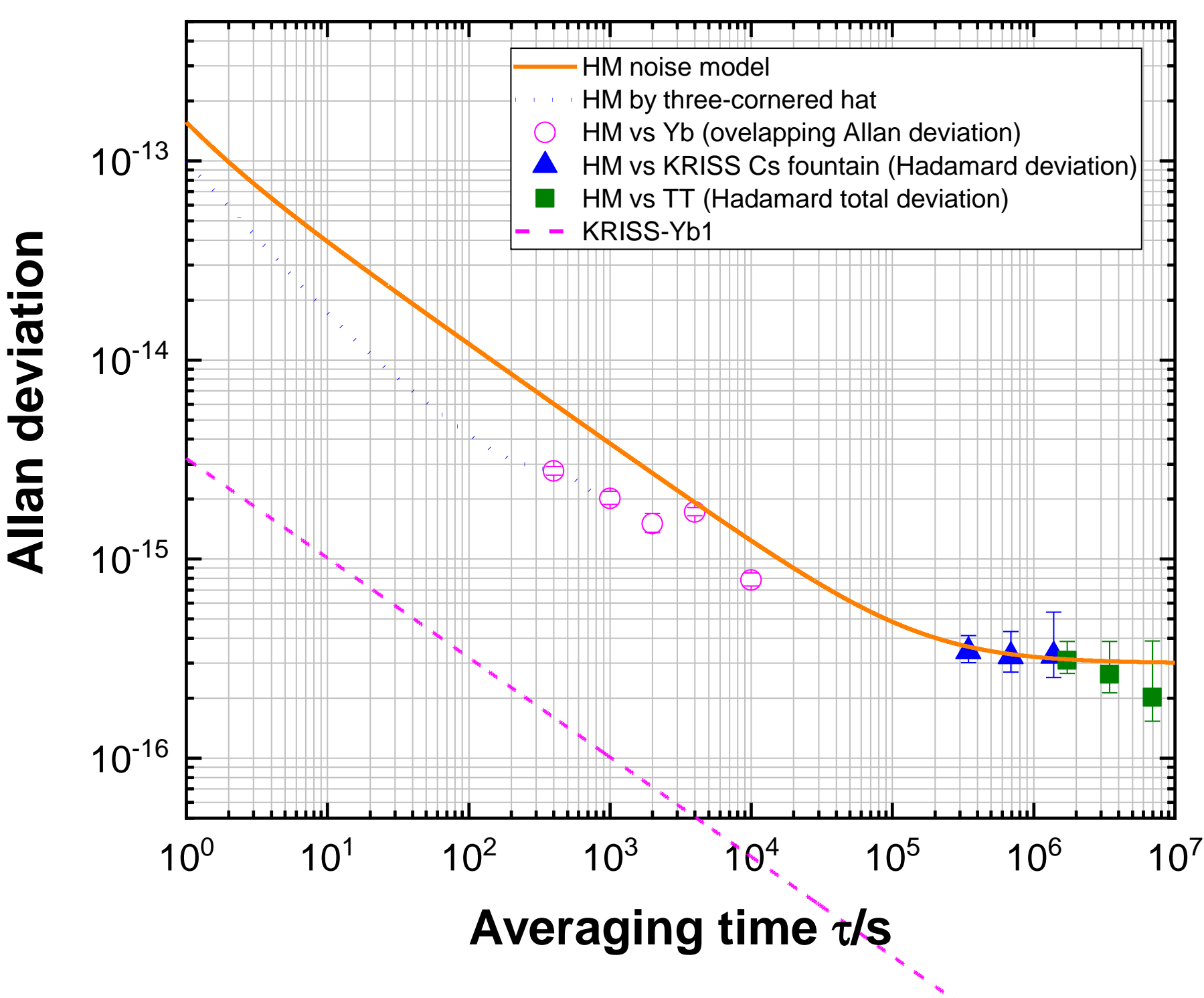

**Figure 7.** Frequency instability of the HM (H5626) measured by three-cornered hat method using two other HMs (blue dotted line), by the Yb optical lattice clock (pink open circles), by the KRISS Cs fountain clock (blue triangles), by TT (green squares). The pink dashed line represents the stability of the Yb optical lattice clock.

### 2.4. UTC(KRIS), Satellite link, TAI, and TT

UTC(KRIS) is the local representation of Coordinated Universal Time (UTC) and the national time and frequency standard in Korea, and it is currently generated by using a HM and a micro-phase stepper (AOG-110, Microsemi) for steering to UTC. UTC(KRIS) is compared using GPSPPP (GPS precise point positioning) and TWSTFT (two-way satellite time and frequency transfer). The comparison data are reported to the BIPM every day and contribute to the 5-day-grid computation of TAI. The frequency deviation of TAI from TT (the SI second on the geoid) is determined by the data of primary and secondary frequency standards (PSFS) worldwide reported in Circular T. During the 14-month-long measurement in this research, 13 primary frequency standards reported data to the BIPM including the Cs fountain clocks (IT-CsF2 [60], METAS-FOC2 [61], NIM5 [62], NPL-CsF2 [63], NRC-FCs2 [4], PTB-CSF1, PTB-CSF2 [2], SU-CsFO2 [64], SYRTE-FO1, SYRTE-FO2, SYRTE-FOM [65]) and two thermal beam Cs clocks (PTB-CS1 and PTB-CS2 [66]). A Rb fountain clock (SYRTYE-FORb [67]) also contributed to TAI. Two optical clocks contributed to TAI during this campaign (NICT-Sr1 [68] (MJD 58914~58934) and NMIJ-Yb1 [26] (MJD 58754~59194)).

## 3. Data analysis – frequency shifts and uncertainties

The frequency shifts and their uncertainties in each step of the absolute frequency measurement are analyzed in this section.

### 3.1. Yb systematic uncertainty

We re-evaluated the systematic uncertainties of the KRISS-Yb1 after several updates in the optical clock system since 2017. The result of the newly evaluated systematic shifts and uncertainty values are summarized in table 1. The total systematic uncertainty of KRISS-Yb1 was $1.7\times10^{-17}$.

**Table 1.** Uncertainty budget for the $^{171}$Yb optical clock (KRISS-Yb1)

| Effect | Relative shift ($\times 10^{-17}$) | Relative Uncertainty ($\times 10^{-17}$) |
|---|---|---|
| Density shift | -0.3 | 0.1 |
| Lattice Stark shift | 1.3 | 0.8 |
| Quadratic Zeeman shift | -6.7 | 0.1 |
| Blackbody radiation | -232.2 | 1.0 |
| DC Stark shift | 0.0 | 1.0 |
| Background gas shift | -1.3 | 0.1 |
| Probe light shift | 0.1 | 0.03 |
| Servo error | 0.0 | 0.3 |
| Line pulling | 0.0 | 0.1 |
| AOM chirp | 0.0 | 0.3 |
| Yb Total | -239.2 | 1.7 |

The trap conditions were changed compared to our previous measurement in 2017 because the sideband cooling and the trap depth control have been adopted as shown in figure 4, thus, the desnity shift was re-evaluated. Although the density shift was at low $10^{-18}$ level as we operated the Yb clock at low atomic density utilizing a large beam waist of 100 μm, a precise density shift evaluation is an essential prerequisite for other uncertainty evaluations if variations of those effects are accompanied by density change, as is in the case of the lattice ac Stark shift. We assumed the density shift to be proportional to the trapped atom number ($N_{atom}$). We obtained the dependence of the density shift on $N_{atom}$ by numerical curve fits at various conditions of lattice trap. $N_{atom}$ was controlled by varying the intensity of the first cooling laser (399 nm) during the core-shell MOT. We calculated the trap depth ($U_0$), the effective potential depth ($U_e$), and the axial motional quantum number ($\overline{n}$) from the sideband spectrum [57]. The other parameters, which affected the density shift, including the mean excitation fraction [57], the pulse area of Rabi spectroscopy [15], and the second MOT compression condition, were kept approximately constant during this measurement campaign. For the typical condition of the absolute frequency measurement ($U_e$ = 28(2) $E_R$, $U_0$=48 $E_R$), the slope of the linear fitting

was –0.74(27) μHz/$N_{\mathrm{atom}}$. The typical density shift was estimated to be -0.3(1)$\times 10^{-17}$ with 2,300(520) atoms trapped in the optical lattice on average, as shown in table 1.

The ac Stark shift of the lattice beam was modeled as

$$\Delta\nu_{\mathrm{ls}} = (a\Delta\nu - b)\left(\bar{n} + \frac{1}{2}\right)\left(\frac{U_e}{E_R}\right)^{1/2} - \left(a\Delta\nu + \frac{3}{4}d(2\bar{n}^2 + 2\bar{n} + 1)\right)\left(\frac{U_e}{E_R}\right)$$
$$+d(2\bar{n} + 1)\left(\frac{U_e}{E_R}\right)^{3/2} - d\left(\frac{U_e}{E_R}\right)^2, \quad (2)$$

where $\Delta\nu_{\mathrm{ls}}$ is the frequency shift by the lattice light, *a* is the slope of the differential electric dipole (E1) polarizability, $\Delta\nu$ is the lattice frequency detuning from the E1 magic frequency. The coefficients *b* and *d* come from the multipolar polarizability and the hyperpolarizability, respectively [58, 69]. For the uncertainty evaluation of the ac Stark shift for our KRISS-Yb1 system, we measured the E1 magic frequency and the coefficients *a* following the analysis introduced in [58]. We did not used a simplified model in [70], which requires that the longitudinal temperature is linearly proportional to the trap depth, because this condition did not match our system. To suppress the residual ac Stark shift from the ASE of the lattice laser [71], we used a ring cavity Ti:sapphire laser instead of a TA, and adopted a narrow bandpass filter. The power build-up cavity for the lattice beam also reduced the effect of the ASE. The E1 magic frequency and the coefficient *a* were determined to be 394 798 257.1(18) MHz and 22.3(19) μHz/MHz, respectively, using the measurements interleaving $U_{\mathrm{e}}$ between 27 $E_{\mathrm{R}}$ and 385 $E_{\mathrm{R}}$. The density shift should be taken into account in the lattice shift evaluation for each different potential depth as described in the last paragraph. For example, for $U_{\mathrm{e}}$ =385 $E_{\mathrm{R}}$, we utilized the experimentally determined value of –0.21(16) μHz/$N_{\mathrm{atom}}$. We adopted the values for the coefficients *b* and *d* from the report of RIKEN [58] for the evaluation of the nonlinear ac Stark shift in this paper. The typical ac Stark shift was given by 1.3(8)$\times 10^{-17}$ for the entire campaign. The typical trap depth $U_0$ was 48 $E_R$ except for the initial two months (January and March 2020), when we used the trap depth of 160 $E_R$ .

We note that there are disagreements in the reported parameter values in equation (2) among research groups including our result [17, 28, 30, 38, 58, 70]. With *a* and E1 magic frequency that were measured by ourselves, we adopted the RIKEN values for *b* and *d* because our experimental condition was more similar to the RIKEN case than others. But we checked whether there would be a difference in our uncertainty evaluation result if we used other parameter values. We tried estimating the lattice shift using two other sets of parameter values. The first trial utilized all the parameters of *a*, *b*, *d,* and E1 magic frequency from the values reported by RIKEN [58], and the second trial utilized all the parameter values reported by INRIM [30]. The calculated values were different from our results by less than $4\times10^{-18}$, which is within the uncertainty of the lattice shift evaluation. In our experiment, the lattice light shift was relatively small and insensitive to the parameter values, mainly because we operated the Yb clock at low trap depth (48 $E_R$).

We followed the analysis of our paper in 2017 [28] to calculate the frequency shift due to the black body radiation (BBR) with the improved thermal environmental condition around the Yb spectroscopy chamber. We adopted the values of the differential static polarizability and the dynamic correction used in [72]. The temperature data from ten calibrated PT100 sensors were logged every 10 seconds. The typical temperature value of the Yb spectroscopy chamber was (22.4 ± 0.3) °C. The temperature uncertainty was conservatively estimated to be half of the difference between the maximum and minimum values during each clock operation time. The uncertainty of the BBR shift from the ambient room temperature radiation through the vacuum chamber windows was reduced compared to that reported in 2017 by a more precise measurement of the transmittance of the AR-coated fused silica windows of the Yb spectroscopy vacuum chamber. We kept the temperature of the coolant that flowed around the AH coils at 15°C using a commercial recirculating chiller for a more homogeneous temperature distribution. To maintain the temperature of the spectroscopy chamber approximately the same and to reduce the waiting time for the thermal equilibrium before the clock operation, we applied an appropriate current to the AH coils during the time the optical clock was not operated. The temperature of the Yb spectroscopy chamber in an idle state was maintained within 0.5°C compared to the temperature during the clock operation by this procedure. The average BBR shift was estimated to be $-232.2(10)\times10^{-17}$.

The coefficient of the quadratic Zeeman shift was measured to be $-1.529(30)$ μHz/Hz$^2$ by varying the strength of the quantization field. During the clock operation, the first order Zeeman splitting was maintained to be 150 Hz from the center with a typical uncertainty of 1.3 Hz. Thus, the quadratic Zeeman shift was evaluated to be $-6.7(1)\times10^{-17}$

The servo error results from an uncompensated residual drift of the clock laser during the interrogation of the clock operation. To evaluate the frequency shift from the servo error, the error signal from the PID servo used for the correction to the center of the clock transition was averaged during the whole campaign to be -1.64(28) mHz. We estimated conservatively the servo error uncertainty to be $3\times10^{-18}$ using the square sum of the average servo error offset and its uncertainty.

The collision between Yb atom and background gases induces a shift of the clock transition frequency. The background gas was dominated by $H_2$, which was determined by a residual gas analyzer. This shift was reported to be proportional to the trap loss rate with a coefficient of $-1.64(12)\times10^{-17}$ s [6]. The frequency shift from background gas collisions was estimated to be $-1.3(1)\times10^{-17}$ using the measured trap lifetime of 1.14(5) s.

The line pulling offset comes from the unwanted spectrums including the lattice sidebands and other clock transitions (π or σ) other than the targeted π transition. The effects from the axial lattice sidebands and other clock transitions were negligible ($< 10^{-19}$). However, the radial lattice sidebands caused the asymmetry to the lineshape of the carrier, which induced a slight offset of the center of frequency. This offset frequency was estimated to be 0.6 mHz, which is taken as an uncertainty of $1\times10^{-18}$.

For the AOM chirp effect and the DC Stark effect, we followed the results of the paper in 2017 [28] because there had been no changes in the experimental system related to these effects. The height of Yb atoms in the optical lattice from the conventionally adopted geoid potential (62 636 856.0 $m^2/s^2$) was newly measured to be 74.66(9) m. The height from the earth ellipsoid (WGS84) [73] was measured by a GPS antenna and orthometric height levelling. The geoid undulation was obtained by a regional geoid model based on the Earth Gravitational Model (EGM2008) [74]. The gravitational shift was conservatively estimated to be $81.4(3)\times10^{-16}$, considering the tidal effect.

### 3.2. Yb clock relative to the HM by the optical frequency comb; $\boldsymbol{f(\mathrm{Yb})_{T(up)}/f(\mathrm{HM})_{T(up)}}$ and $\boldsymbol{f(\mathrm{HM})_{T(up)}/f(\mathrm{HM})_{T(5d)}}$

The frequency of the Yb optical lattice clock relative to the HM was measured using an optical frequency comb and a frequency counter described in Section 2.2. The measurement was performed from January 2020 to February 2021 with a total measurement time of 400.5 h and a total uptime of 5.7%. The clock operation of the whole system was so robust that the continuous frequency measurement could be as long as 111,410 s, and monthly uptime could be as high as 15% in March 2020. The major cause for the interruption of the optical clock operation was the failure of the frequency stabilization of the lattice laser. We measured the absolute frequency intermittently with the typical duration of a single run between 8,000 s and 12,000 s. As the barycenter of the measurement data was not coincident with the midpoint of the 5-day grid of the Circular T, the drift compensation of the HM was required. The drift rate of the HM and its uncertainty for each month-grid of the Circular T (figure 6) were determined by a linear fit of the frequency of H5626 relative to TT for 3 months (including the previous and the next month). The frequency at the barycenter of the data was determined by a linear fit of the data points of measurement runs obtained in each 5-day grid with a fixed slope given by the HM drift rate, and with a pivot point at the barycenter. The y-intercept error of the fit at the pivot point was considered to be the statistical error in determining the frequency at the barycenter. The weighting factor of each data point was given by the inverse square of the uncertainty. Next, we estimated the frequency at the midpoint of each 5-day grid by using the monthly determined HM drift rate. We considered an additional statistical uncertainty in this process using the drift rate uncertainty and the time difference between the barycenter and the midpoint.

The short-term frequency stability of the H5626 was estimated to be $1\times10^{-13}$ at 1 s by the three-cornered-hat method using two other HM as shown in figure 7. However, the comb-measured instability of the frequency comparison of the Yb clock and the HM was $2\sim3\times10^{-13}$ at 1 s due to additional comb noise, which decreased fast as $1/\tau$ to around 10 s. The three-cornered hat measurement and the comb measurement agree with each other after 100 s, as shown in figure 7. During the time interval when we used H5628 as a frequency reference, we processed the measured data to be relative to H5626 by using the DMTD, which continuously measured the phase difference between the two HMs. The resulting short-term instability was slightly larger in this case ($5\sim6\times10^{-13}$ at 1 s) due to the measurement stability limit of the DMTD. We conservatively took the statistical uncertainty of each run to be the overlapping Allan deviation at the end point of the comb measurement (typically at a quarter of the measurement duration).

The dead-time uncertainty due to the intermittent measurement in each 5-day-long interval was estimated following the procedure presented in [75] with the HM noise model shown in Section 2.3 and figure 7.

The systematic uncertainty of the frequency conversion by the optical frequency comb from RF to optical frequency was estimated to be $7\times10^{-17}$ by comparing the simultaneous measurement of the 1156 nm clock laser by two similar optical frequency combs.

To summarize the measurement uncertainties up to this point, the systematic uncertainty in the link between the Yb clock and H5626 ($\mathrm{u}_B/\mathrm{lab}$) is given by the systematic uncertainty from the frequency comb. The statistical uncertainty in the link between the Yb clock and H5626 ($\mathrm{u}_A/\mathrm{lab}$) is given by the square root of the square sum of the statistical uncertainties from the frequency comb measurement, the HM drift compensation and the dead-time of the Yb clock measurement.

### 3.3. $f(\mathrm{HM})_{T(5d)}/f(\mathrm{UTC(KRIS)})_{T(5d)}$ and uncertainty

$f(\mathrm{HM})_{T(5d)}/f(\mathrm{UTC(KRIS)})_{T(5d)}$ for each 5-day grid was measured by the DMTD, which continuously measured the phase difference between the HM and UTC(KRIS) without dead-time. The DMTD phase measurement data were obtained at 1 s interval with the inaccuracy and the resolution well below 0.01 ns. Thus, the statistical uncertainty and the systematic uncertainty of the $f(\mathrm{HM})_{T(5d)}/f(\mathrm{UTC(KRIS)})_{T(5d)}$ measurement was less than $1\times10^{-17}$ and negligible for the absolute frequency measurement.

### 3.4. Satellite link, $f(\mathrm{UTC(KRIS)})_{T(5d)}/f(\mathrm{TAI})_{T(5d)}$, and uncertainty

The frequency ratio $f(\mathrm{UTC(KRIS)})_{T(5d)}/f(\mathrm{TAI})_{T(5d)}$ was calculated using the phase difference of UTC(KRIS) and TAI is published in Circular T every month in 5-day grid. UTC(KRIS) was compared with TAI by GNSS satellite link. The uncertainty in the link to TAI is given by [76]

$$u_{l/TAI} = \frac{\sqrt{2}u_{link}}{T_0} / \left(\frac{T}{T_0}\right)^{0.9},$$

where $T$ is the length of the measurement time, $T_0 = 5$ d, and $u_{link} = 0.3$ ns is the statistical (type A) uncertainty of the time transfer from UTC(KRIS) to TAI as published in Circular T.

### 3.5. $f(\mathrm{TAI})_{T(5d)}/f(\mathrm{TAI})_{T(m)}$, $f(\mathrm{TAI})_{T(m)}/f(\mathrm{SI})_{T(m)}$, and uncertainties

The monthly frequency deviation of TAI from the SI second (TT), $f(\mathrm{TAI})_{T(m)}/f(\mathrm{SI})_{T(m)}$ is obtained using the "duration of the TAI scale interval *d*" reported in Circular T taking the opposite sign. We should extrapolate the 5-day grid data (relative to TAI) to the corresponding month interval (25, 30, or 35 days depending on the month). The drift of $f(\mathrm{TAI})/f(\mathrm{SI})$ was negligible, but the dead-time uncertainty of $f(\mathrm{TAI})_{T(5d)}/f(\mathrm{TAI})_{T(m)}$ was estimated following the procedure presented in [75] using the frequency instability of flywheel EAL (free atomic time scale), which is explained in the document "Explanatory supplement to BIPM Circular T" [77] and is published for each month at BIPM FTP site [78]. For Circular T 385~396, the EAL instability was the square sum of white frequency noise $1.4\times10^{-15}$ $(\tau/\mathrm{d})^{-1/2}$, and flicker frequency noise $3.0\times10^{-16}$, and random walk frequency noise $2.0\times10^{-17}$ $(\tau/\mathrm{d})^{1/2}$.

Thereafter, for Circular T 397~398, flicker frequency noise was $2.0 \times 10^{-16}$, and white frequency noise and random walk frequency noise were the same.

### 3.6. Overall uncertainty and correlation

The absolute frequency values of KRISS-Yb1 were obtained for each 5-day grid during the 14-month-long measurement campaign with the uncertainties combining all the factors described in this section.

Next, the absolute frequency for each month corresponding to a Circular T report was obtained by the weighted mean of the 5-day-grid measurements available in that period. And then the final absolute frequency value for the total measurement campaign was determined by the weighted mean of the monthly obtained values. The weights were given by the combined uncertainties for each measurement. The result for each month is summarized in table 2. The uptime for each month and the relative frequency offset from the 2017 CIPM-recommended frequency of 518 295 836 590 863.6(3) Hz [5] are also reported in table 2.

The correlations between the data sets were considered following the procedure in [79] using covariance matrixes. The Yb systematic uncertainties, the frequency comb systematic uncertainties, the gravitational redshift uncertainties were considered to be fully correlated in time. If any primary or secondary frequency standard was reported in more than two Circular T's, the systematic uncertainty of that was considered to be fully correlated.

The statistical uncertainties of the HM drift estimation, which utilized three-month-long measurements relative to TT, were considered to be partially correlated if the measurement period overlapped. The adjacent values of $f(\mathrm{UTC(KRIS)})_{T(5d)}/f(\mathrm{TAI})_{T(5d)}$ share the same phase difference data reported in Circular T either as the initial phase or the final phase; thus, the corresponding statistical uncertainties of the satellite transfer for adjacent 5-grid measurements were considered to have a negative correlation. The other statistical uncertainties were considered to be uncorrelated in time.

**Table 2.** Uncertainty budget of the KRISS-Yb1 absolute frequency measurement relative to the SI second for each month and for the total measurement campaign. Yb/SI represents the relative frequency offset from the 2017 CIPM-recommended frequency of 518 295 836 590 863.6(3) Hz. The uncertainty and the frequency offset are given by the factor of $10^{-16}$ of relative frequency.

| | Cir.T 385 | Cir.T 387 | Cir.T 389 | Cir.T 390 | Cir.T 391 | Cir.T 392 | Cir.T 393 | Cir.T 394 | Cir.T 395 | Cir.T 397 | Cir.T 398 | **Total** |
|---|---|---|---|---|---|---|---|---|---|---|---|---|
| Uptime(%) | 1.7 | 15.0 | 1.6 | 5.9 | 7.9 | 5.4 | 2.9 | 1.6 | 9.1 | 8.0 | 1.8 | **5.7** |
| **Yb/SI** | 0.4 | -4.3 | 13.6 | 1.8 | 2.3 | 4.1 | 0.9 | 16.0 | -5.8 | 12.5 | -1.1 | **2.9** |
| Type A uncertainty | | | | | | | | | | | | |
| Yb/HM | 4.9 | 2.5 | 7.1 | 4.2 | 3.0 | 4.2 | 5.9 | 6.4 | 3.3 | 2.9 | 5.1 | **1.2** |
| HM drift | 0.17 | 0.06 | 0.06 | 0.13 | 0.11 | 0.16 | 0.09 | 0.17 | 0.11 | 0.26 | 0.05 | **0.07** |
| HM dead time | 6.3 | 2.7 | 6.3 | 3.8 | 3.5 | 4.3 | 5.1 | 6.8 | 3.1 | 3.5 | 6.5 | **1.3** |
| Satellite link | 4.9 | 3.2 | 9.8 | 2.9 | 5.0 | 4.4 | 4.5 | 5.0 | 4.8 | 2.3 | 8.5 | **1.2** |
| TAI dead time | 4.1 | 3.0 | 6.4 | 0.0 | 3.0 | 2.7 | 1.8 | 4.4 | 2.4 | 1.3 | 4.7 | **0.8** |
| PSFS | 0.9 | 0.5 | 0.6 | 0.6 | 1.2 | 0.7 | 0.7 | 0.5 | 0.6 | 0.6 | 0.7 | **0.2** |
| **Total type A** | **10.3** | **5.7** | **15.1** | **6.4** | **7.5** | **8.0** | **9.2** | **11.5** | **7.1** | **5.3** | **12.8** | **2.3** |
| Type B uncertainty | | | | | | | | | | | | |
| Yb | 0.29 | 0.23 | 0.19 | 0.16 | 0.15 | 0.15 | 0.14 | 0.14 | 0.14 | 0.14 | 0.15 | **0.2** |
| Frequency comb | 0.7 | 0.7 | 0.7 | 0.7 | 0.7 | 0.7 | 0.7 | 0.7 | 0.7 | 0.7 | 0.7 | **0.7** |
| Gravitational shift | 0.3 | 0.3 | 0.3 | 0.3 | 0.3 | 0.3 | 0.3 | 0.3 | 0.3 | 0.3 | 0.3 | **0.3** |
| PSFS | 1.2 | 1.2 | 1.1 | 1.2 | 0.9 | 1.2 | 1.2 | 1.2 | 1.0 | 1.0 | 1.1 | **1.0** |
| **Total type B** | **1.4** | **1.4** | **1.4** | **1.4** | **1.2** | **1.4** | **1.4** | **1.4** | **1.3** | **1.3** | **1.3** | **1.2** |
| **Total uncertainty** | **10.4** | **5.9** | **15.2** | **6.5** | **7.6** | **8.1** | **9.3** | **11.6** | **7.2** | **5.5** | **12.9** | **2.6** |

### 3.7. Summary of 5-day-grid Yb frequency measurements relative to TAI

The frequency measurement values of KRISS-Yb1 relative to TAI in 5-day grids are shown in figure 8. The monthly determined KRISS-Yb1 frequencies relative to TAI are also shown with black dotted lines. This result agrees well with the duration of the TAI scale interval *d*, which represents the frequency deviation of TAI from the SI second by taking the opposite sign. This result implies that KRISS-Yb1 can be used as a good secondary frequency standard for TAI calibration in Circular T.

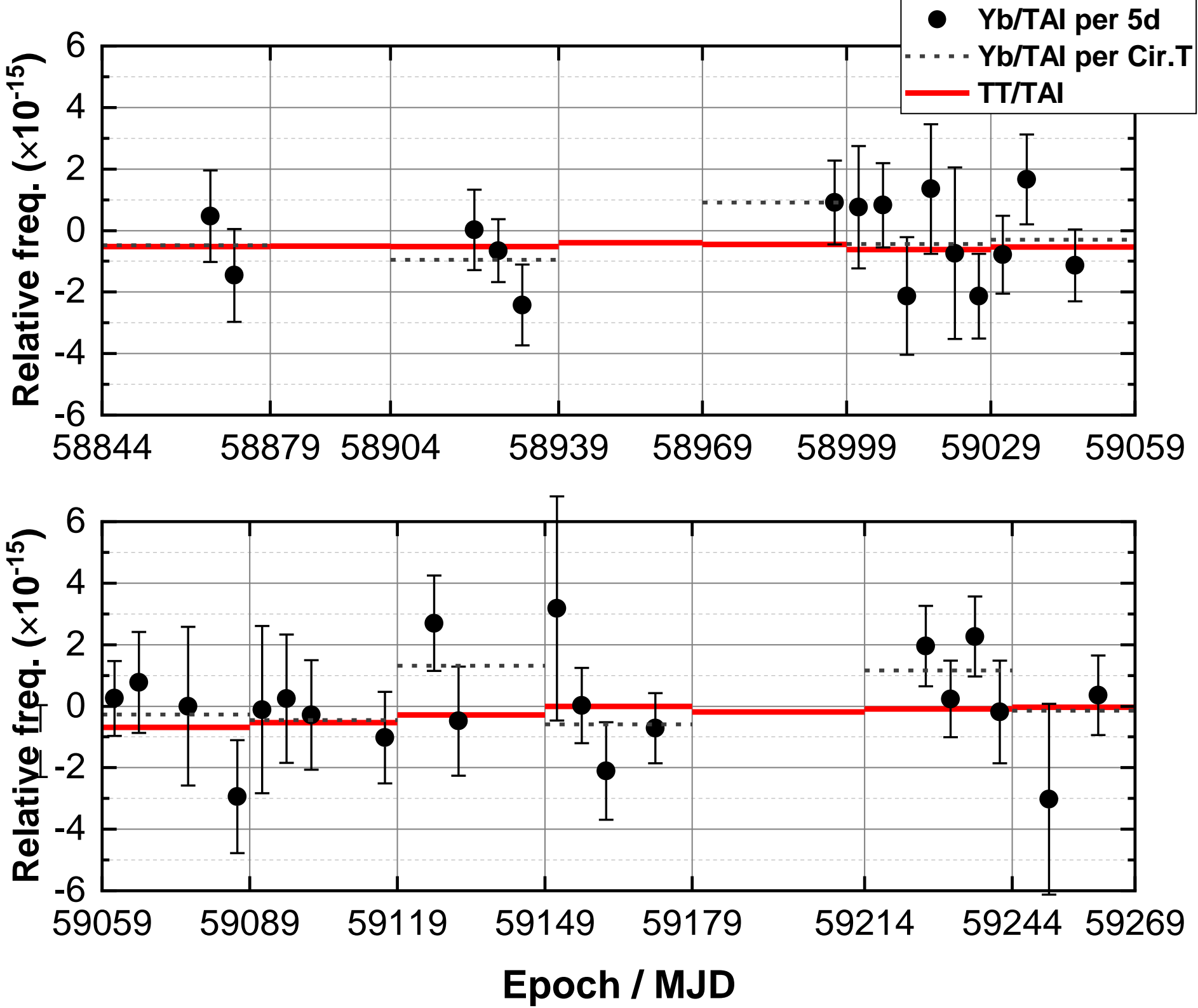

**Figure 8.** The relative frequency offset from the 2017 CIPM-recommended frequency of the 5-day-grid KRISS-Yb1 measurement values relative to TAI (black dots), the monthly determined KRISS-Yb1 frequency (black dashed lines), and the duration of the TAI scale interval *d* (TT relative to TAI) reported in Circular T (thick red lines).

## 4. Result

The absolute frequencies for the months corresponding to each Circular T are shown in figure 9. The total uncertainties (blue error bars) for each month were between $5\times10^{-16}$ and $1.5\times10^{-15}$ depending on the uptime of the KRISS-Yb1. The systematic uncertainties (light green error bars) were between $1.2\times10^{-16}$ and $1.5\times10^{-16}$. The final absolute frequency in the total measurement campaign was determined to be 518 295 836 590 863.75(14) Hz with the relative frequency uncertainty of $2.6\times10^{-16}$ by the weighted mean of the values for each month (red line), and it agrees well with the recommended value of CIPM in 2017 (gray shade).

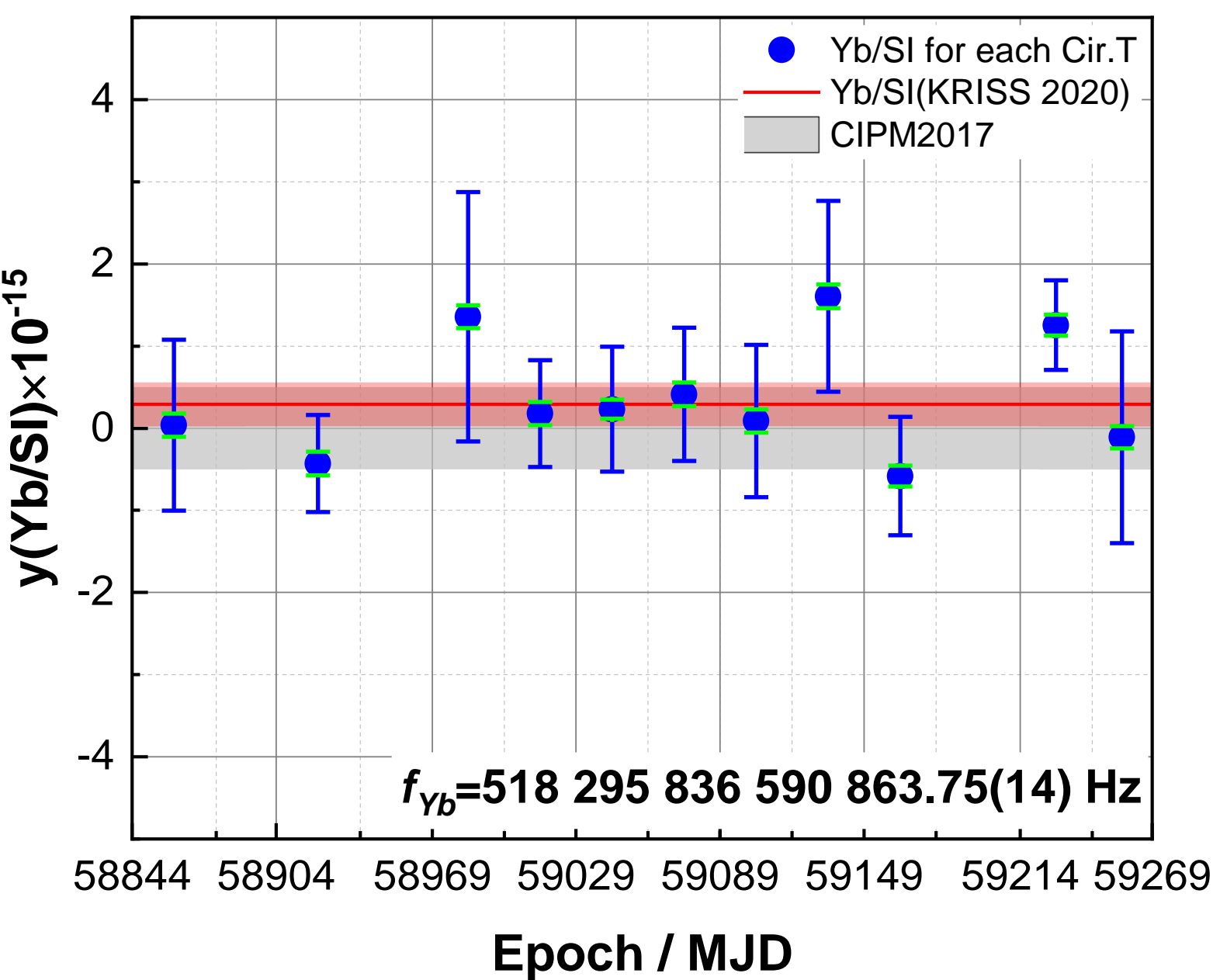

**Figure 9.** The relative frequency offset of the monthly determined KRISS-Yb1 absolute frequencies for each Circular T (blue dots) from the 2017 CIPM-recommended frequency. Blue error bars represent the total uncertainties of each month and light green error bars represent

the type B uncertainties. The weighted mean during the total measurement campaign was 518 295 836 590 863.75 Hz (red line) with the total uncertainty of 0.14 Hz (red shade). The uncertainty of the 2017 CIPM recommended frequency is shown as a gray shade.

The $^{171}$Yb absolute frequency measurement in this campaign is compared with other results by different laboratories worldwide, which is summarized in figure 10 including the results relative to TAI, relative to a local Cs standard, and from the frequency ratio between optical clocks of different atomic species. The specific experimental methods adopted in various measurements are described in the figure caption. The year of the publication in a peer-reviewed journal was taken to be the time when each measurement was performed. As can be seen, the measurement results at KRISS agree well with the other measurements and CIPM recommendation frequency.

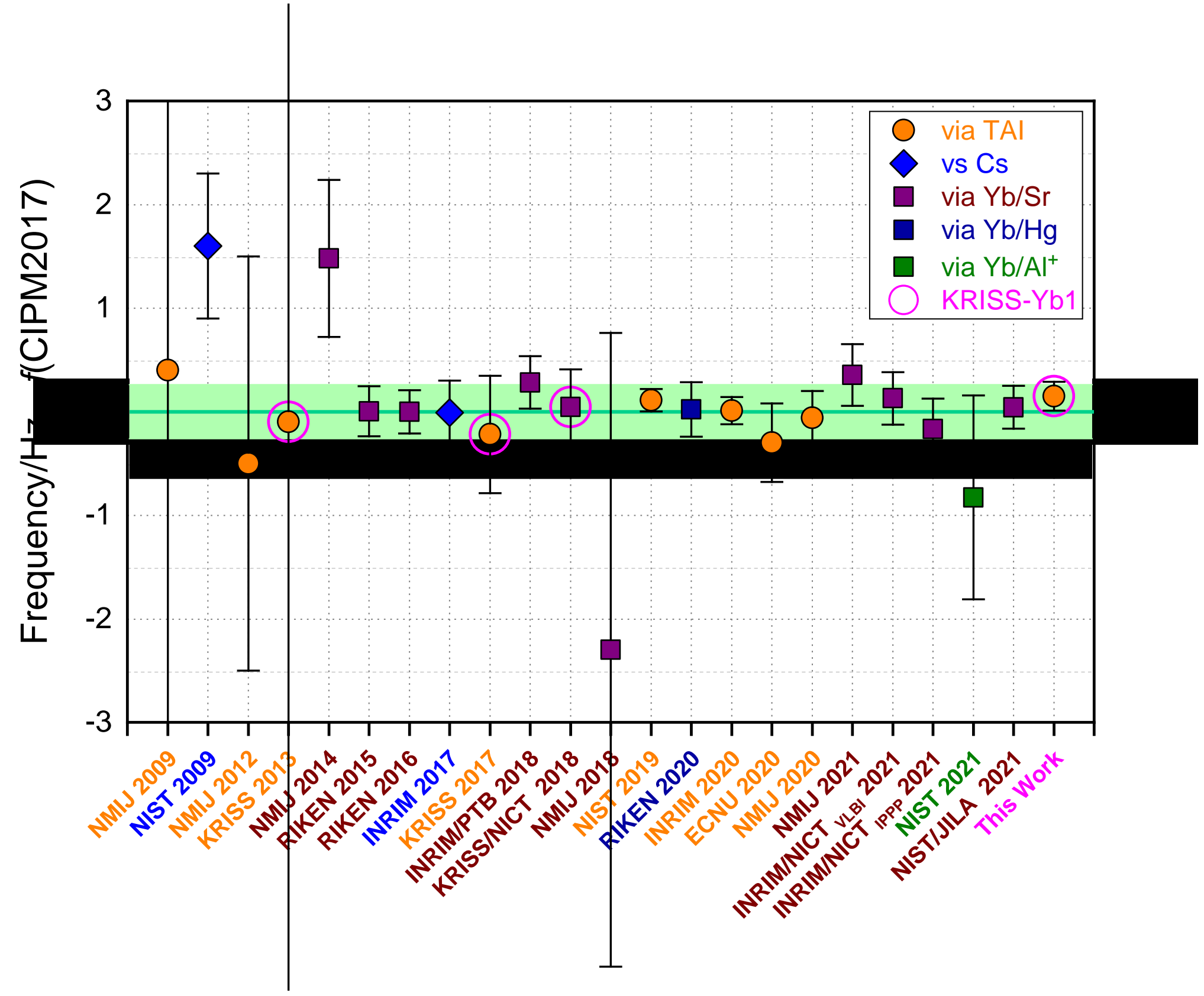

**Figure 10.** History of the $^{171}$Yb absolute frequency measurement by laboratories worldwide relative to TAI (orange circles including the results from NMIJ [13, 25, 26], KRISS [27, 28], NIST [29], INRIM [30], and ECNU [31])), relative to a local Cs fountain clock (blue diamonds including the results from NIST [32] and INRIM [20]). Also, the calculated absolute frequency values using the frequency ratio of optical clocks with different atomic species and the CIPM recommended frequency [5] are shown together by the Yb/Sr ratio (purple squares including the results from NMIJ [33-36], RIKEN [17, 37], INRIM/PTB [39], KRISS/NICT [40], INRIM/NICT [41] and NIST/JILA [42], by the Yb/Hg ratio from RIKEN [38] (navy blue square), and by the Yb/Al$^+$ ratio from NIST [42] (green square). The frequency ratio measurements were performed by various methods, such as local clock comparisons [17, 33-38], a transportable optical clock [39], an advanced satellite-based frequency transfer (TWCP) [40], very long baseline interferometry (VLBI) [41], an optical fiber link [42], and a free-space link [42]. The uncertainty of the CIPM recommended frequency (0.3 Hz) is shown with a green shade. The measurements at KRISS are shown with hollow pink circles, which agrees well with the other measurements and with the CIPM-recommended frequency.

## 5. Conclusion

We performed the absolute frequency measurement of the $^{1}S_0$-$^{3}P_0$ transition of a $^{171}$Yb optical lattice clock (KRISS-Yb1) for 14 months relative to the SI second (TT). The result agrees well with the CIPM-recommended frequency and with recent measurements performed by other laboratories with uncertainties at low $10^{-16}$. This result is expected to contribute to the future update of the CIPM recommendation frequency. Also, this agrees well with the TAI calibration results by primary and secondary frequency standards worldwide. Thus, it can be operated as one of the good optical secondary frequency standards, which will contribute to the calibration of TAI and to the future redefinition of the second.

## Acknowledgements

The authors thank Sung-Hoon Yang, Young Kyu Lee, Sang-Wook Hwang, Joon Hyo Lee, Jong Koo Lee, Ho Seong Lee, and Chang Bok Lee for the UTC(KRIS) frequency comparison using satellites and for the measurement of the GNSS ellipsoidal height, Sang Eon Park for the operation of the KRISS Cs fountain clock, Taeg Yong Kwon for providing the DMTD frequency data, Young Hee Lee for the calibration of the temperature sensors, and Ki-Lyong Jeong for the transmission measurement of the vacuum view-ports. We also thank the research groups worldwide that operated primary and secondary frequency standards reported in Circular T. This work was supported by Research on Measurement Standards for Redefinition of SI Units funded by Korea Research Institute of Standards and Science (KRISS – 2021 – GP2021-0001).

# Corrigendum: Absolute frequency measurement of the $^{171}$Yb optical lattice clock at KRISS using TAI for over a year (2021 *Metrologia* 58 055007 [1])

Huidong Kim, Myoung-Sun Heo, Chang Yong Park, Dai-Hyuk Yu, and Won-Kyu Lee*

Korea Research Institute of Standards and Science, Daejeon 34113, South Korea

*E-mail: oneqlee@kriss.re.kr

**Abstract**

In this corrigendum, we correct the height of Yb atoms in the optical lattice clock to be 75.15(4) m based on the new height measurement [2]. Accordingly, the gravitational shift of 81.4(3) $\times 10^{-16}$ is corrected to 81.9(3) $\times 10^{-16}$, and the determined absolute frequency, 518 295 836 590 863.75(14) Hz, is corrected to 518 295 836 590 863.73(14) Hz.

---

**in Abstract**

The determined absolute frequency is corrected to 518 295 836 590 863.73(14) Hz.

### 3.1. Yb systematic uncertainty

The height of Yb atoms in the optical lattice from the conventionally adopted geoid potential (62 636 856.0 $m^2/s^2$) should be corrected to be 75.15(4) m. The gravitational shift should be corrected to be 81.9(3)$\times 10^{-16}$.

**Table 2.** (Fourth row (Yb/SI) should be corrected as below considering the new height measurement [2]) Uncertainty budget of the KRISS-Yb1 absolute frequency measurement relative to the SI second for each month and for the total measurement campaign. Yb/SI represents the relative frequency offset from the 2017 CIPM-recommended frequency of 518 295 836 590 863.6(3) Hz. The uncertainty and the frequency offset are given by the factor of $10^{-16}$ of relative frequency.

| | Cir.T 385 | Cir.T 387 | Cir.T 389 | Cir.T 390 | Cir.T 391 | Cir.T 392 | Cir.T 393 | Cir.T 394 | Cir.T 395 | Cir.T 397 | Cir.T 398 | **Total** |
|---|---|---|---|---|---|---|---|---|---|---|---|---|
| Uptime(%) | 1.7 | 15.0 | 1.6 | 5.9 | 7.9 | 5.4 | 2.9 | 1.6 | 9.1 | 8.0 | 1.8 | **5.7** |
| **Yb/SI** | -0.1 | -4.8 | 13.1 | 1.3 | 1.8 | 3.6 | 0.4 | 15.5 | -6.3 | 12.0 | -1.6 | **2.4** |
| Type A uncertainty | | | | | | | | | | | | |
| Yb/HM | 4.9 | 2.5 | 7.1 | 4.2 | 3.0 | 4.2 | 5.9 | 6.4 | 3.3 | 2.9 | 5.1 | **1.2** |
| HM drift | 0.17 | 0.06 | 0.06 | 0.13 | 0.11 | 0.16 | 0.09 | 0.17 | 0.11 | 0.26 | 0.05 | **0.07** |
| HM dead time | 6.3 | 2.7 | 6.3 | 3.8 | 3.5 | 4.3 | 5.1 | 6.8 | 3.1 | 3.5 | 6.5 | **1.3** |
| Satellite link | 4.9 | 3.2 | 9.8 | 2.9 | 5.0 | 4.4 | 4.5 | 5.0 | 4.8 | 2.3 | 8.5 | **1.2** |
| TAI dead time | 4.1 | 3.0 | 6.4 | 0.0 | 3.0 | 2.7 | 1.8 | 4.4 | 2.4 | 1.3 | 4.7 | **0.8** |
| PSFS | 0.9 | 0.5 | 0.6 | 0.6 | 1.2 | 0.7 | 0.7 | 0.5 | 0.6 | 0.6 | 0.7 | **0.2** |
| **Total type A** | **10.3** | **5.7** | **15.1** | **6.4** | **7.5** | **8.0** | **9.2** | **11.5** | **7.1** | **5.3** | **12.8** | **2.3** |
| Type B uncertainty | | | | | | | | | | | | |
| Yb | 0.29 | 0.23 | 0.19 | 0.16 | 0.15 | 0.15 | 0.14 | 0.14 | 0.14 | 0.14 | 0.15 | **0.2** |
| Frequency comb | 0.7 | 0.7 | 0.7 | 0.7 | 0.7 | 0.7 | 0.7 | 0.7 | 0.7 | 0.7 | 0.7 | **0.7** |
| Gravitational shift | 0.3 | 0.3 | 0.3 | 0.3 | 0.3 | 0.3 | 0.3 | 0.3 | 0.3 | 0.3 | 0.3 | **0.3** |
| PSFS | 1.2 | 1.2 | 1.1 | 1.2 | 0.9 | 1.2 | 1.2 | 1.2 | 1.0 | 1.0 | 1.1 | **1.0** |
| **Total type B** | **1.4** | **1.4** | **1.4** | **1.4** | **1.2** | **1.4** | **1.4** | **1.4** | **1.3** | **1.3** | **1.3** | **1.2** |
| **Total uncertainty** | **10.4** | **5.9** | **15.2** | **6.5** | **7.6** | **8.1** | **9.3** | **11.6** | **7.2** | **5.5** | **12.9** | **2.6** |

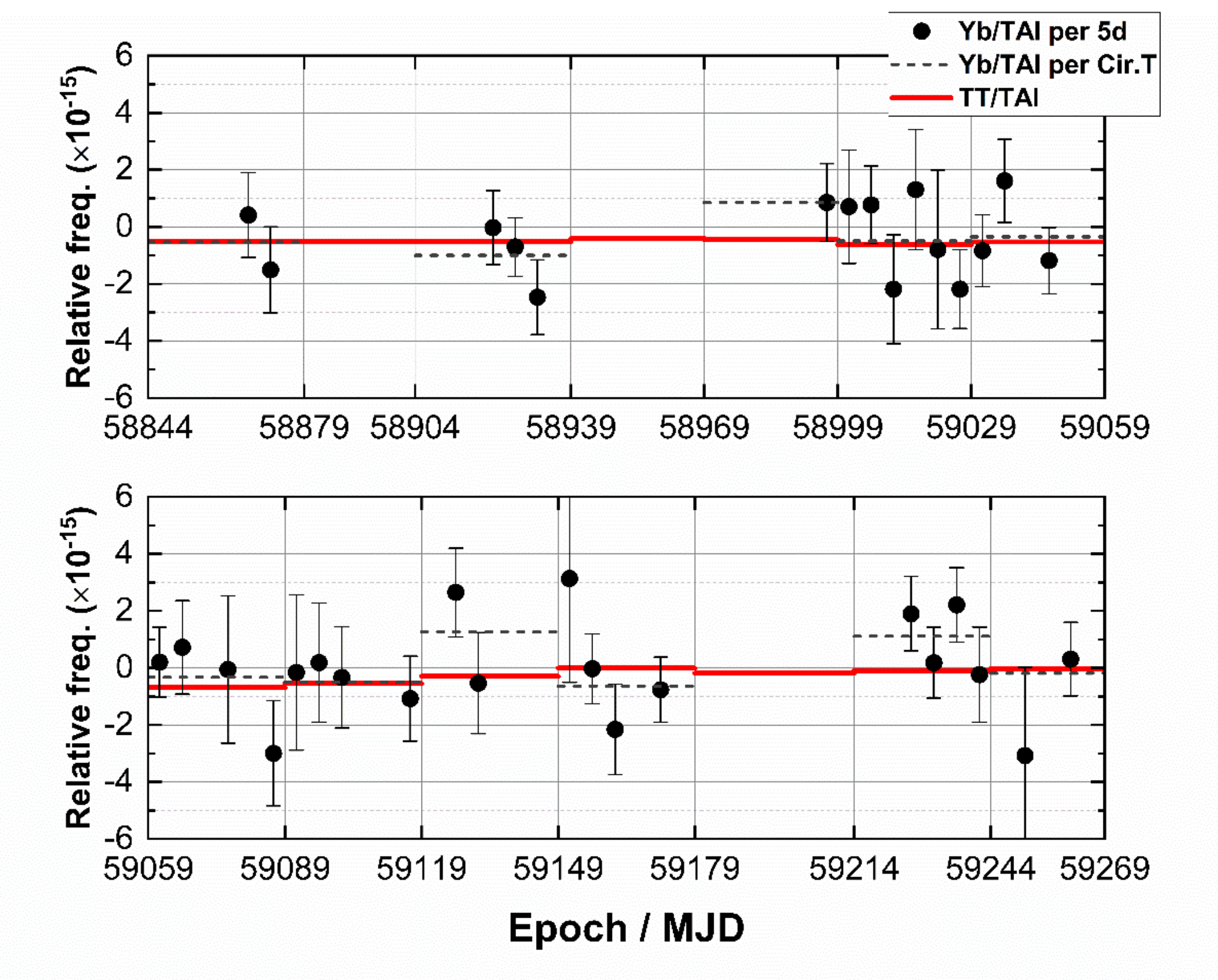

**Figure 8.** (Yb/TAI points are redrawn considering the new height measurement [2]) The relative frequency offset from the 2017 CIPM-recommended frequency of the 5-day-grid KRISS-Yb1 measurement values relative to TAI (black dots), the monthly determined KRISS-Yb1 frequency (black dashed lines), and the duration of the TAI scale interval *d* (TT relative to TAI) reported in Circular T (thick red lines).

**4.Result**

The final absolute frequency in the total measurement campaign should be corrected to be 518 295 836 590 863.73(14) Hz.

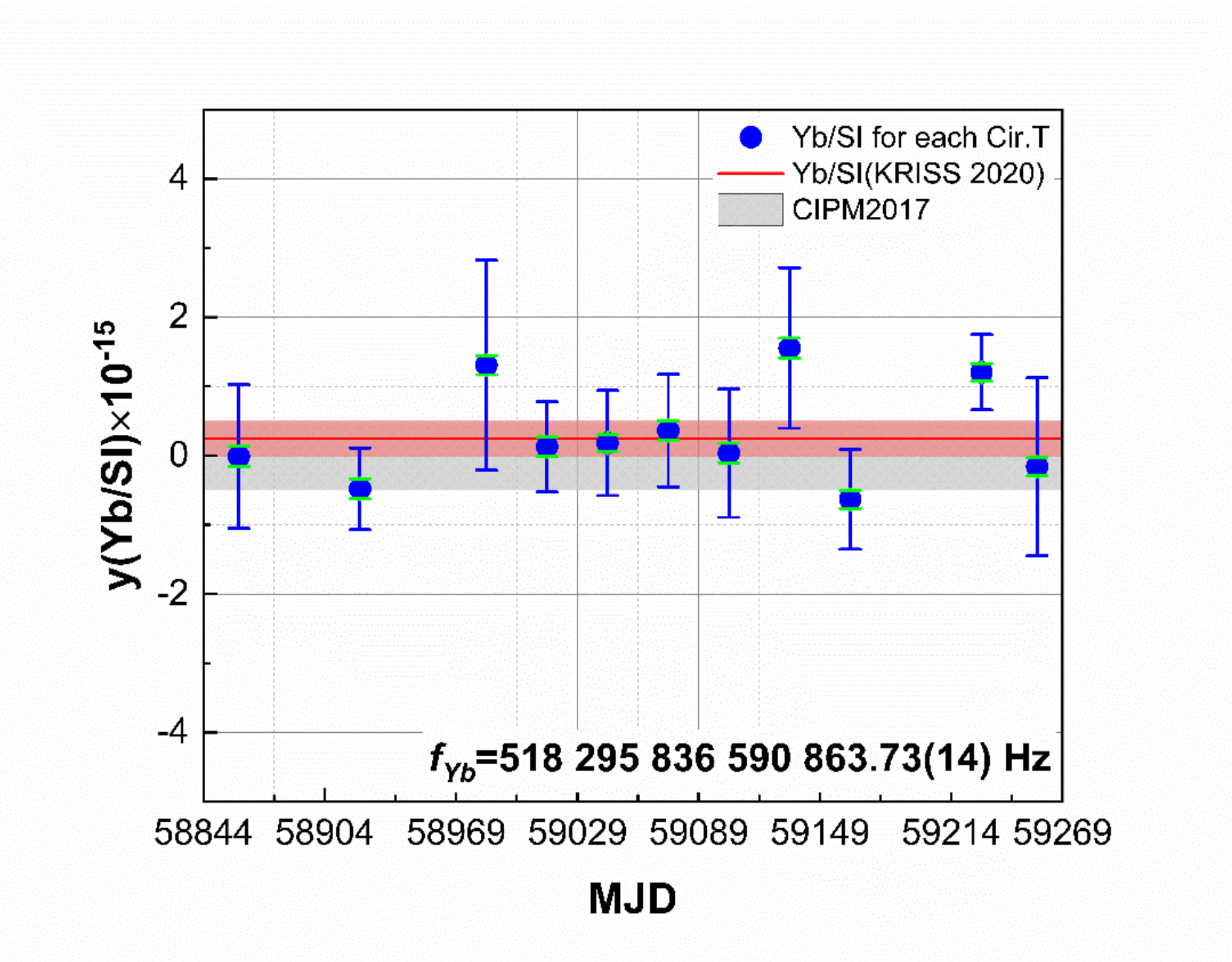

**Figure 9.** (Yb/SI points are redrawn considering the new height measurement [2]) The relative frequency offset of the monthly determined KRISS-Yb1 absolute frequencies for each Circular T (blue dots) from the 2017 CIPM-recommended frequency. Blue error bars represent the total uncertainties of each month and light green error bars represent the type B uncertainties. The weighted mean during the total measurement campaign was 518 295 836 590 863.73 Hz (red line) with the total uncertainty of 0.14 Hz (red shade). The uncertainty of the 2017 CIPM recommended frequency is shown as a gray shade.

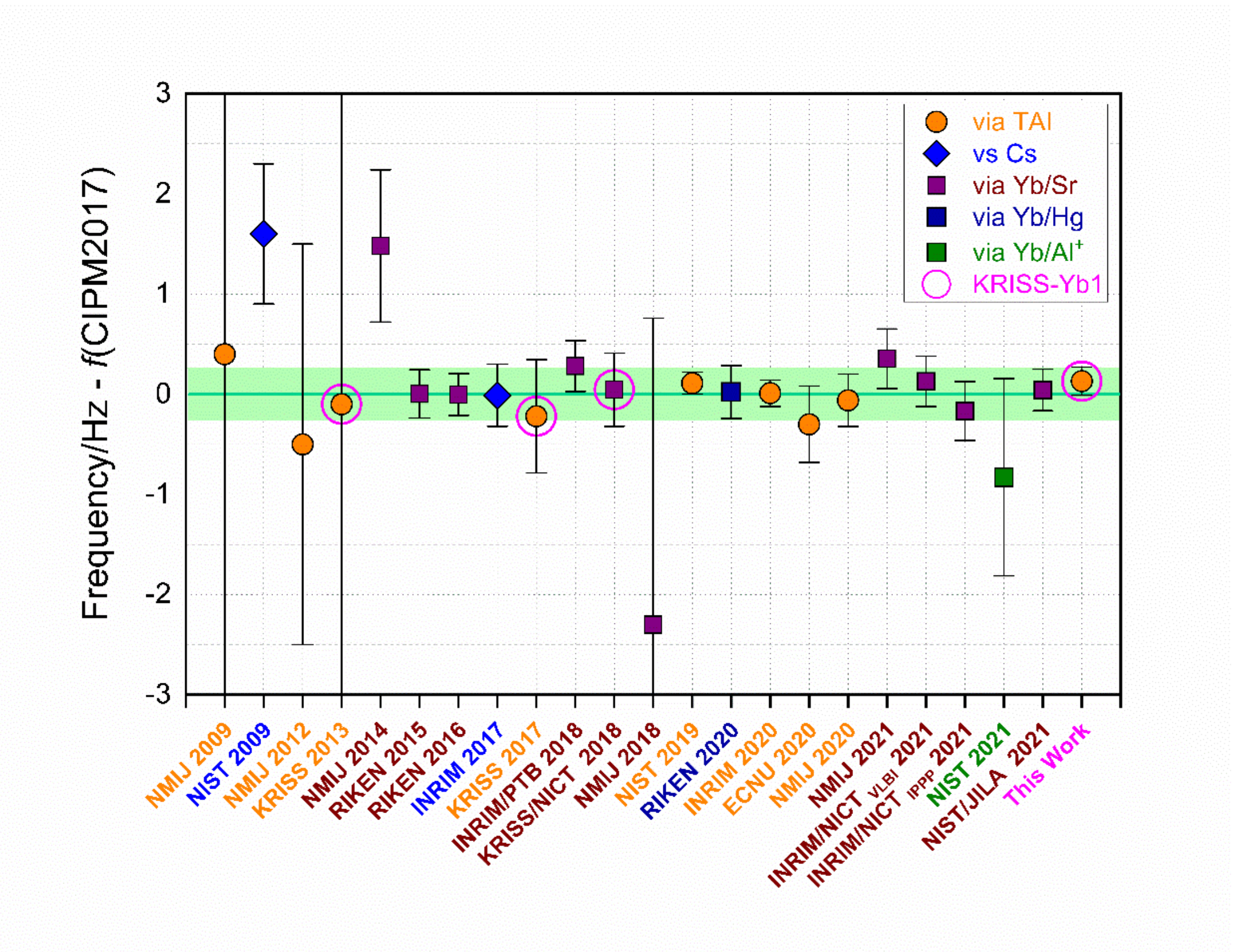

**Figure 10.** (The last measurement point (This Work) is redrawn considering the new height measurement [2])